\documentclass{PoS}
\usepackage{amssymb}
\usepackage{bm}

\newcommand{\be}{\begin{equation}}
\newcommand{\ee}{\end{equation}}
\newcommand\beq{\begin{eqnarray}}
\newcommand\eeq{\end{eqnarray}} 
\newcommand\eqn[1]{\label{eq:#1}} 
\newcommand\eq[1]{eq. (\ref{eq:#1})}

\newcommand{\bfq}{{\mathbf q}}

\newcommand{\MeV}{{\rm ~MeV }}

\newcommand{\bfsigma}{\boldsymbol{ \sigma}}
\newcommand{\bftau}{\boldsymbol{ \tau}}

\title{New approach to NN with perturbative pions}

\ShortTitle{New approach to NN with perturbative pions}

\author{\speaker{Silas Beane}%
\\
        University of New Hampshire \\
        E-mail: \email{silas@physics.unh.edu}}

\abstract{The current status of effective field theory (EFT)
descriptions of nucleon-nucleon (NN) interactions is briefly reviewed,
and a new formulation of EFT which treats pion interactions
perturbatively is presented. This approach differs from the
Kaplan-Savage-Wise (KSW) expansion in that the singular short distance
part of the pion tensor interaction is summed to all orders.}

\FullConference{6th International Workshop on Chiral Dynamics\\
                 July 6-10 2009\\
                 Bern, Switzerland}

\begin{document}

\section{Weinberg and KSW}

\noindent Weinberg was the first to describe an EFT for nuclear forces
\cite{Weinberg:1990rz,Weinberg:1991um,Weinberg:1992yk}, and devised
the prescription that one compute the nuclear potential in an EFT
expansion, truncate at a given order, and then solve the
Lippmann-Schwinger equation exactly with that potential.  This program
has since been pursued by a number of groups
\cite{Ordonez:1992xp,Ordonez:1993tn,vanKolck:1994yi,Ordonez:1995rz,Friar:1998zt,Rentmeester:1999vw,
Bernard:1996gq,Epelbaum:1999dj,Epelbaum:2000mx,Epelbaum:2002ji,Epelbaum:2003xx,Epelbaum:2004fk,
Entem:2001cg,Entem:2002sf,Entem:2003ft,
PavonValderrama:2005wv,PavonValderrama:2005uj}, with very impressive
fits to phase shift data at ${N}^3{LO}$.  An advantage of this
approach is that the long distance part of the interaction correctly
incorporates chiral symmetry; furthermore, with Weinberg's power
counting scheme for the EFT expansion, there is in principle a
systematic improvement of the results with increasing order.  A
disadvantage of Weinberg's scheme is that in its naive implementation,
there are divergences that cannot be absorbed by operators included at
that order, arising from the singular nature of the EFT potential
\cite{Kaplan:1996xu,Beane:2001bc,Nogga:2005hy}.  Thus results depend
on a regulator scale $\Lambda$ which cannot be removed, implying that
the treatment of short distance interactions is model-dependent.  An
analysis of high partial wave channels at NLO in the Weinberg EFT in
ref. \cite{Nogga:2005hy} demonstrated that the cutoff dependence was a
feature of all channels subject to an attractive pion tensor force ---
despite the fact that there is no local operator to absorb this model
dependence until order $(\ell+1)$ in the expansion for a channel with
angular momentum $\ell$.  Furthermore ref.~\cite{Nogga:2005hy}
demonstrated that at this order, observables in channels with an
attractive tensor interaction are particularly sensitive to the value
of the cutoff even at energies as low as $T_{lab} = 50\MeV$. It is
argued that predictions at a given order only vary at the level of
higher order corrections as the regulator is varied over some range,
so that the model dependence does not interfere with the predictive
power of the EFT.  This hope is difficult to verify since the
computations are all numerical, and the numerical evidence suggests
that the acceptable range for $\Lambda$ is very narrow.

The alternative KSW theory entails an expansion of the NN scattering
amplitude, instead of the nuclear potential, effected by computing a
well-defined class of Feynman diagrams at each order in the expansion
\cite{Kaplan:1996xu, Kaplan:1998we,Kaplan:1998tg}.  KSW power counting
is not determined by how operators scale near the trivial IR fixed
point of the nucleon contact interaction; instead it is determined by
operator scaling about the nontrivial UV fixed point corresponding to
infinite scattering length.  At this fixed point nucleon operators for
s-wave scattering develop large anomalous dimensions and are resummed
nonperturbatively, a reasonable starting point given how much larger
NN scattering lengths are than the range of their interaction.

The KSW scheme expands the NN scattering amplitude in powers of $Q$,
where the nucleon momentum $p$, pion mass $m_\pi$ and the inverse
scattering length $1/a$ are all considered $O(Q)$, while other mass
scales such as the nucleon mass $M$, the pion decay constant $f_\pi$
are taken to be $O(1)$.  It was argued that convergence of the KSW
expansion is governed by the scale $\Lambda_{NN} = 16\pi
f_\pi^2/(g_A^2 M)= 300\MeV$.  An advantage of this approach is that
the scattering amplitudes can be computed analytically, and at each
order the amplitude is renormalized and independent of the cutoff.  NN
phase shifts were computed to order NNLO in
refs. \cite{Fleming:1999bs,Fleming:1999ee}; the result for the
spin-singlet ${}^1S_0$ phase shift is shown in Fig.~1, plotted versus
the momentum $p$ of each nucleon in the center of mass frame.
\begin{figure}[t]
\centerline{\includegraphics[width=8.1cm]{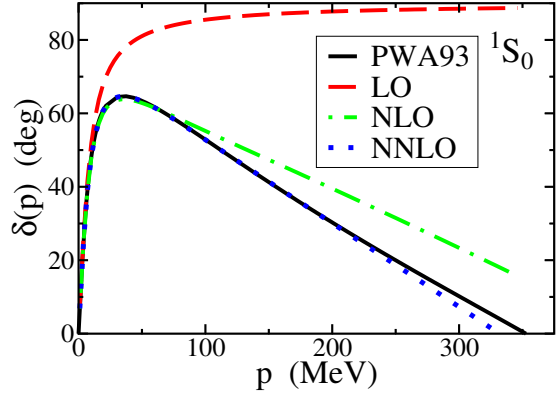}}
\label{J0}
\caption[a]{The ${}^1S_0$ NN phase shift in the KSW expansion, versus
momentum in the center of mass frame to NNLO, compared with the Nijmegen
PWA93 partial wave analysis~\protect\cite{Stoks:1993tb}.}
\end{figure}

Although successful in the spin-singlet channel, it was discovered in
ref.~\cite{Fleming:1999ee} that the KSW expansion does not converge in
the ${}^3S_1$ channel, and the authors identified the singular tensor
potential mediated by pions, scaling as $-1/r^3$ for small $r$, to be
the cause of the failure.  Such a singular attractive interaction is
incapable of supporting a ground state and no contact interaction can
remedy this pathology. One possible solution suggested in
\cite{Beane:2001bc} is to expand around the chiral limit, treating the
infinite number of bound states in the pion potential as being short
range and outside the realm of validity of the EFT. 

\section{A New Expansion}

\noindent In ref.~\cite{Beane:2008bt} we propose a different solution:
we modify the pion propagator in a manner reminiscent of Pauli-Villars
regulation characterized by a heavy mass scale $\lambda$.  This
modification removes the $1/r^3$ singularity in pion exchange,
effectively shifting that physics into the contact interactions and
reordering the summation of strong short-distance effects.  The
advantages of the KSW expansion are retained: there is a well-defined
power counting scheme that organizes the calculation, and results are
analytic.  Dependence on the scale $\lambda$ can therefore be studied
analytically, and we find that all contributions that grow as powers
of $\lambda$ are absorbed into counterterms. The limit $\lambda\to
\infty$ is therefore smooth, and the KSW expansion is recovered in
that limit. Here we present promising results for the low-lying spin
triplet phase shifts to NNLO that indicate convergence of the
expansion.

Our starting point is the assumption that the failure of the KSW
expansion is due to the singular short distance pion tensor
interaction, which can be eliminated by a shift in the contact
interactions of the EFT.  The underlying principles of EFT imply that
we are free to distort the short range pion interactions however we
please, as the counterterms serve to ensure the correct low energy
effects of short distance physics. We therefore choose the
modification in order to: (i) make it possible to analytically perform the
diagrammatic expansion; (ii) leave unaltered the KSW expansion of the
spin-singlet channel, since apparently no convergence
problem is encountered there. These considerations lead us to replace
the pion propagator $G_\pi(\bfq,m_\pi)$ by
 \beq 
 G_\pi(\bfq,m_\pi)
+ G_{(1,1)}(\bfq,\lambda) + G_{(1,0)}(\bfq,\lambda)
 \eqn{gmod} \eeq 
where the subscript $(I,J)$ indicates the isospin and spin of a fictitious
meson. Including couplings at the ends of the propagators, these
expressions are given by
 \beq G_\pi(\bfq,m_\pi) &=& { i \frac{g_A^2}{4
f_\pi^2}\frac{(\bfq\cdot\bfsigma_1)(\bfq\cdot\bfsigma_2)(\bftau_1\cdot\bftau_2)}{\bfq^2
+ m_\pi^2} }\cr G_{(1,0)}(\bfq,\lambda) &=& { i
\frac{g_A^2\lambda^2}{4f_\pi^2}\frac{(\bftau_1\cdot\bftau_2)}{\bfq^2 +
\lambda^2} }\ , \eeq 
and $G_{(1,1)}(\bfq,\lambda) = -G_\pi(\bfq,\lambda)$. The $ G_{(1,1)}
$ term looks like exchange of a pion with the wrong sign propagator
and mass $\lambda$, canceling the short distance $1/r^3$ part of the
pion-induced tensor interaction for $r\lesssim 2\pi/\lambda$. The
$G_{(1,0)}$ term is included to exactly cancel $G_{(1,1)}$ (up to a
contact interaction) in the spin-singlet channel; it resembles the
exchange of an $I=1$, $J=0$ meson, also of mass $\lambda$.  In the
above expressions $g_A\simeq 1.25$ and $f_\pi\simeq 93\MeV$;
${\bfsigma}$ and $\bftau$ are spin and isospin matrices respectively.
Note that the only free parameter is the mass scale $\lambda$.  We
expect that for $\lambda\gtrsim 2 \Lambda_{NN}$ the derivative
expansion is not adversely affected, and that the original KSW
expansion is recovered in the $\lambda\to\infty$ limit.

We emphasize that we are not using $ G_{(1,1)} $ and $ G_{(1,0)}$ to
model real meson exchange, but only as a device to eliminate the
strong short distance behavior from the tensor pion exchange, putting
all that physics in the contact interactions which are fit to data.
Choosing the masses in $ G_{(1,1)} $ and $ G_{(1,0)}$ to both
equal $\lambda$ greatly simplifies the analytic computations.

\section{NNLO calculation of spin-triplet amplitudes}

\noindent Making use of the modified pion propagator \eq{gmod} and
classifying the mass scale $\lambda$ to also be $O(Q)$, we have
computed all the Feynman diagrams in
\cite{Fleming:1999bs,Fleming:1999ee} relevant for the ${}^3S_1$,
${}^3D_1$ and $\epsilon_1$ partial wave channels. These diagrams are
evaluated using dimensional regularization and we choose the
renormalization scale, $\mu=m_\pi$. The analytic formulas for our NNLO
calculations will be given elsewhere; here we present the results
graphically.  In Fig.~2 we show our results with $\lambda=750\MeV$ for
the ${}^3S_1$, ${}^3D_1$ and $\epsilon_1$ phase shifts, compared with
the Nijmegen partial wave analysis \cite{Stoks:1993tb}. All three of
our results are improvements over the NNLO KSW computation in
\cite{Fleming:1999bs,Fleming:1999ee}, and with the exception of
$\epsilon_1$, show signs of converging on the correct answer.  The
result for $\epsilon_1$ is less convincing, but it should be noted
that the anomalously small value for $\epsilon_1$ in nature suggests
that delicate cancellations are at play, and one would only expect an
EFT prediction to start converging at high order in the expansion.
\begin{figure*}[t]
\centerline{\hbox{\includegraphics[width=6.1cm]{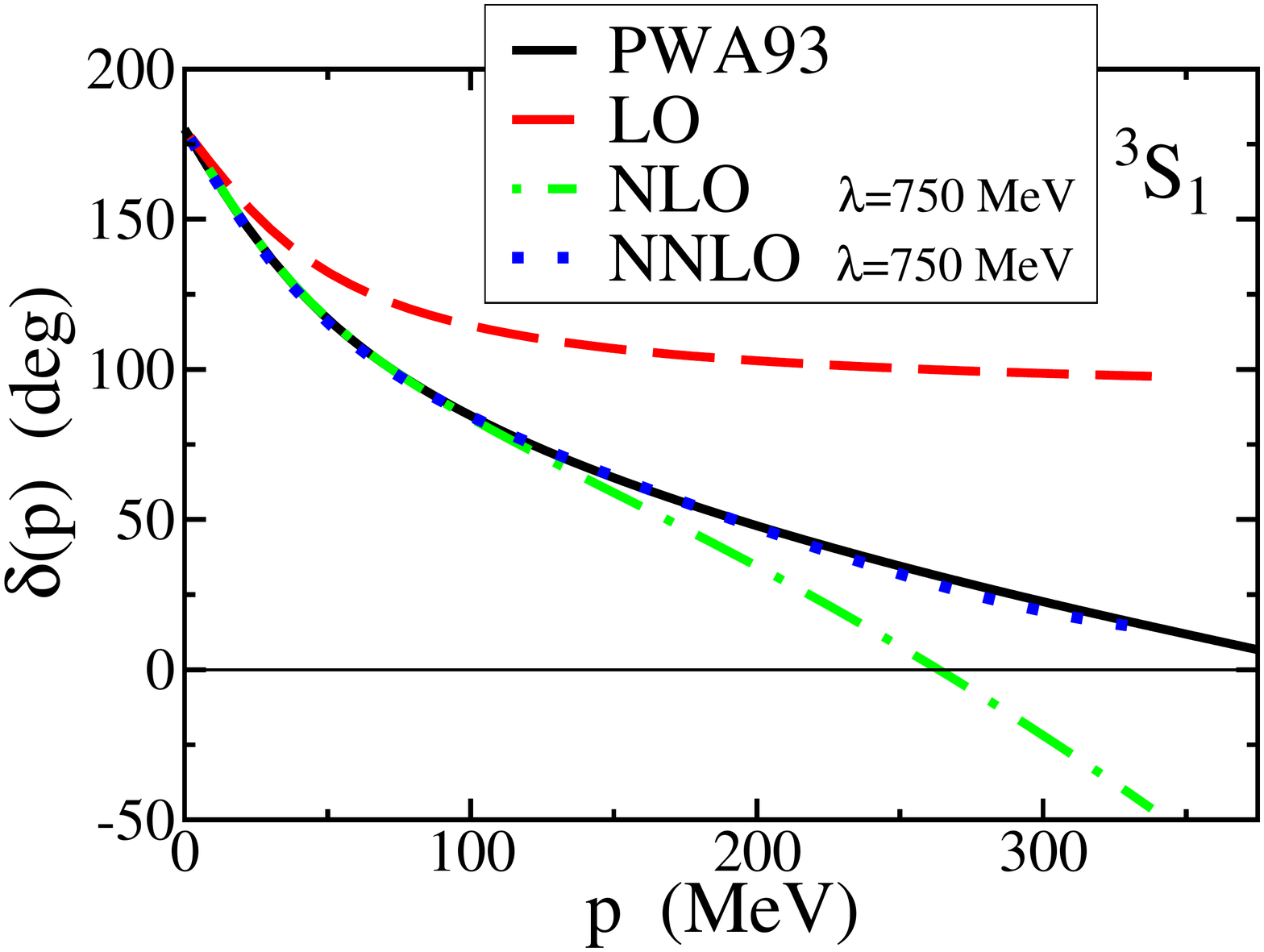}\includegraphics[width=6.1cm]{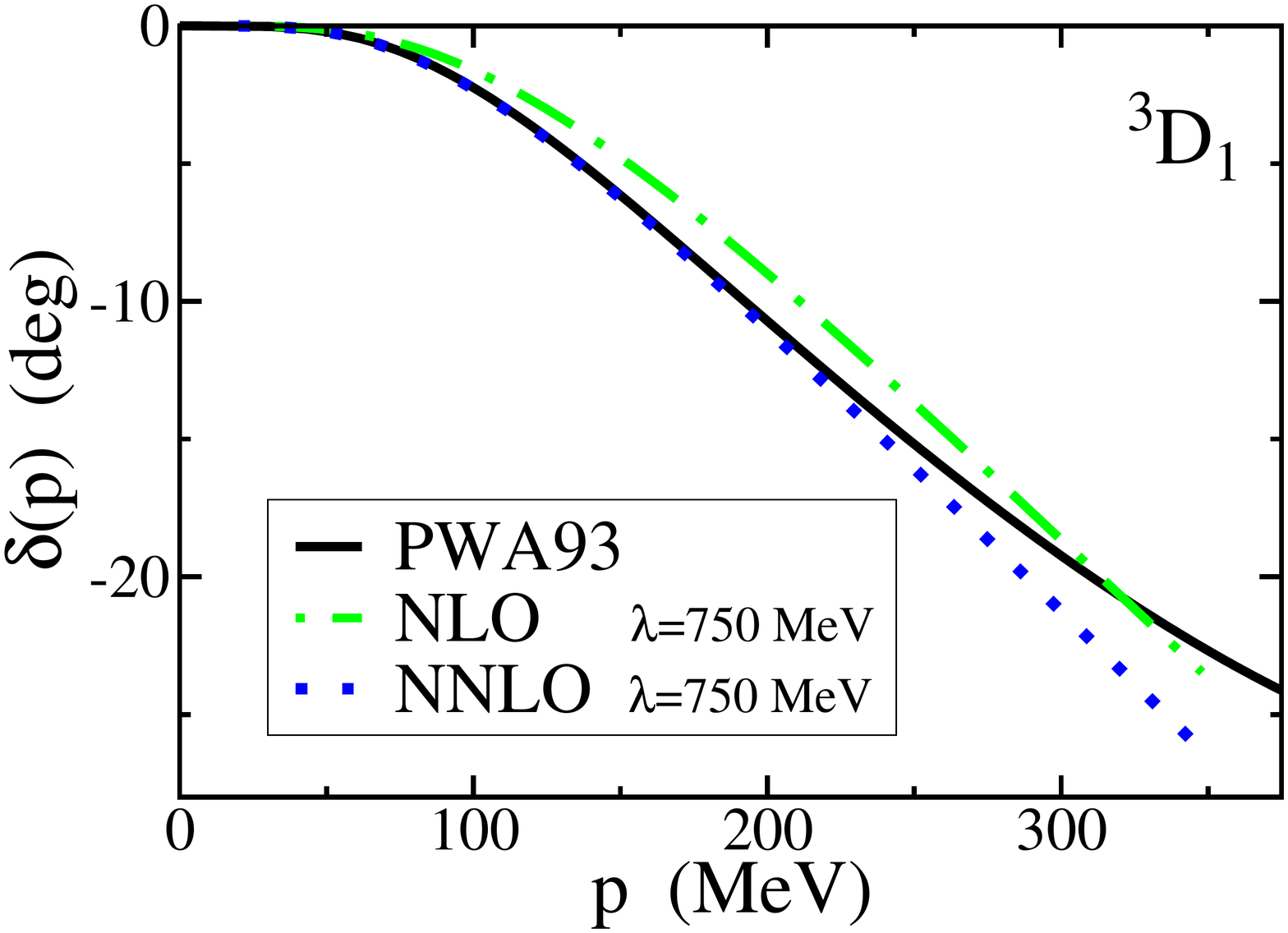}}}
\centerline{\hbox{\includegraphics[width=6.1cm]{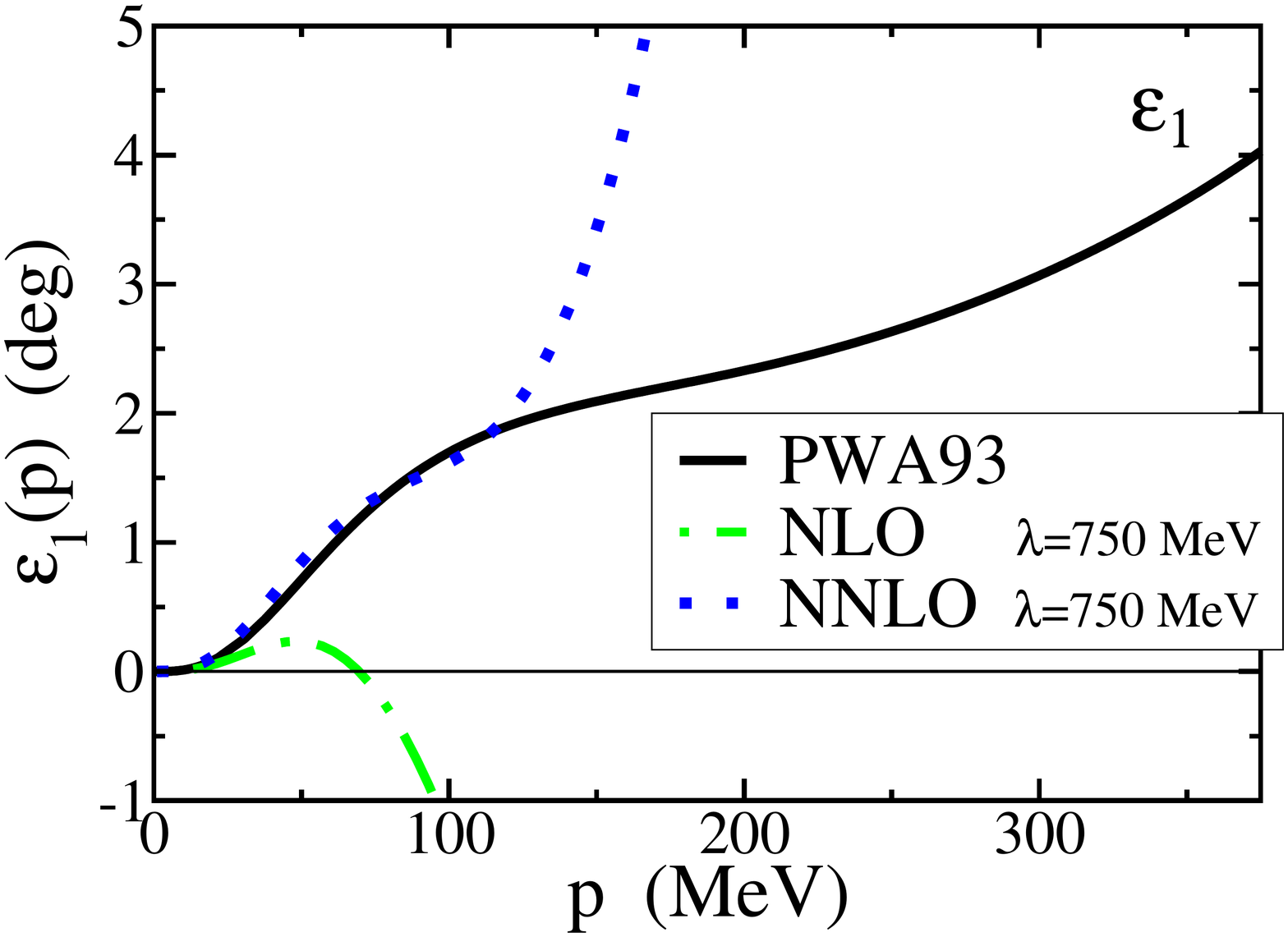}}}
\label{J1gen}
\caption[a]{New results for the ${}^3S_1$, ${}^3D_1$, and
$\bar\epsilon_1 $ phase shifts plotted versus momentum in the center
of mass frame to NNLO, compared with the Nijmegen PWA93 partial wave
analysis.}
\end{figure*}

The dependence of our results on $\lambda$ is displayed in Fig.~3,
where the bands indicate the changes in the phase shifts over the
range $600\MeV\le \lambda\le 1000\MeV$ .  It is apparent from these
figures that our results are not extremely sensitive at low $p$ to the
value we take for $\lambda$.
\begin{figure*}[t]
\centerline{\hbox{\includegraphics[width=6.1cm]{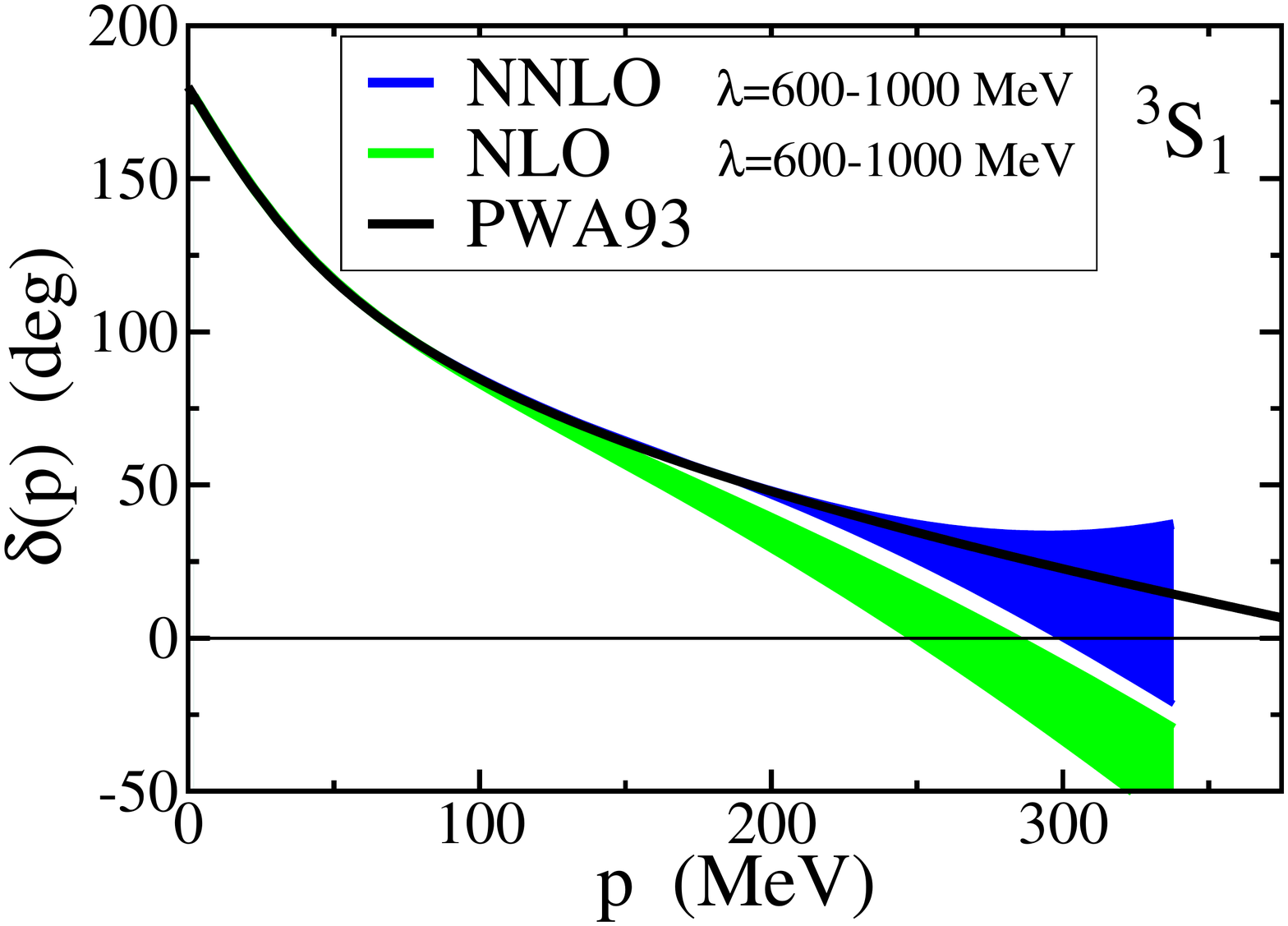}\includegraphics[width=6.1cm]{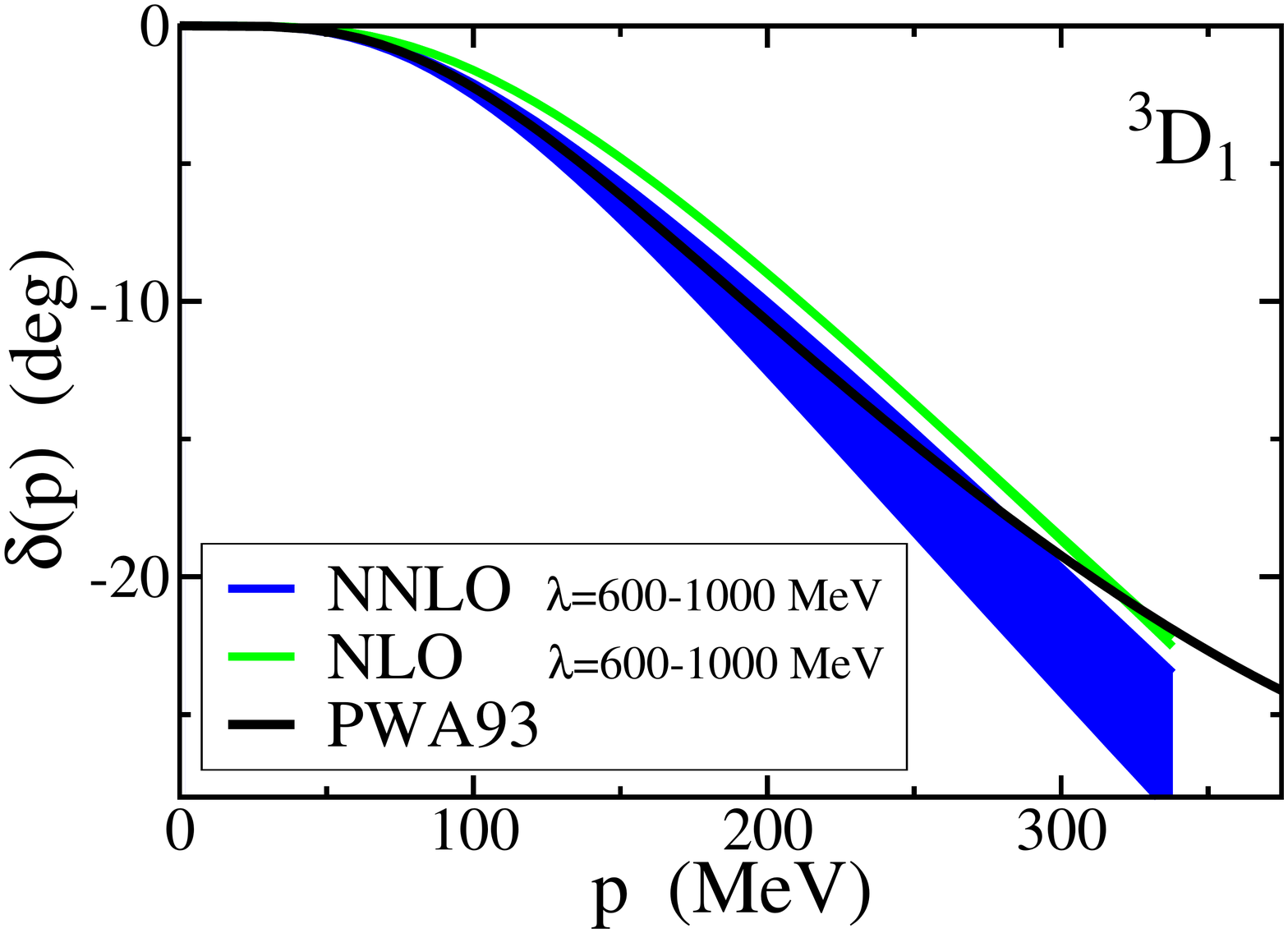}}}
\centerline{\hbox{\includegraphics[width=6.1cm]{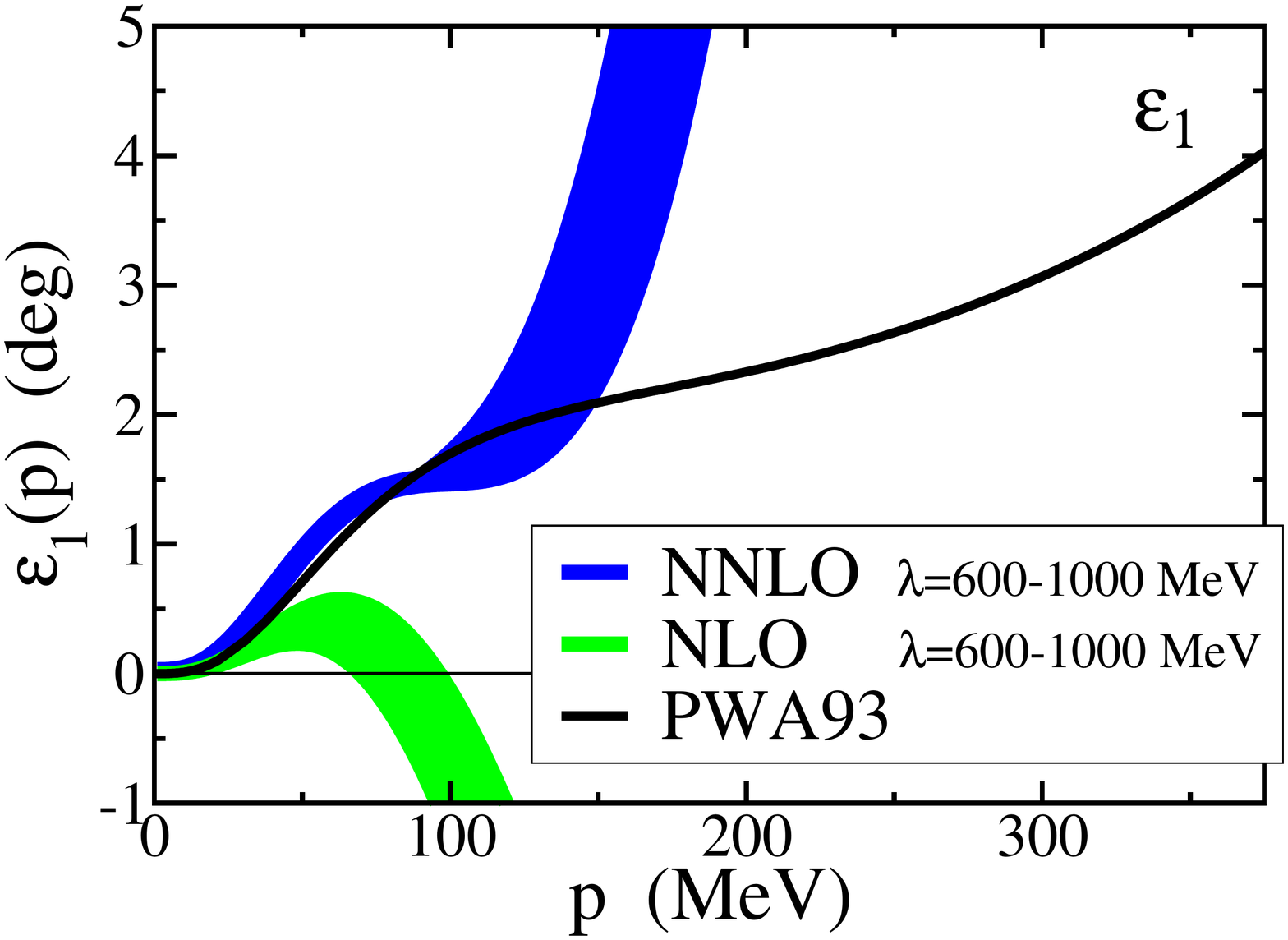}}}
\label{J1var}
\caption[a]{The NLO (green band) and NNLO (blue band) results for the
${}^3S_1$, ${}^3D_1$, and $\bar\epsilon_1 $ phaseshifts showing their
variation as $\lambda$ is varied in the range $600 \MeV \le \lambda
\le 1000 \MeV$.} 
\end{figure*}

The role of the scale $\lambda$ in these calculations can be easily
addressed given the analytic form we have derived for the scattering
amplitudes.  It may seem strange that $\lambda$---a regularization
scale---is being treated as $O(Q)$ which is our low energy expansion
scale.  In particular, one might worry that scattering amplitudes have
terms proportional to powers of $\lambda/\Lambda_{NN}$, which is
formally $O(Q)$ but numerically $>1$.  In fact though one can show
analytically that at each order in the expansion, contributions to the
amplitudes proportional to positive powers of $\lambda$ are all
absorbed into the counterterms available at that order.  Therefore the
amplitudes only depend on inverse powers of $\lambda$, and in the
$\lambda\to \infty$ limit the fictitious meson propagators in
\eq{gmod} decouple and one smoothly recovers the results of
\cite{Fleming:1999bs,Fleming:1999ee}.

\section{Remaining issues}

\noindent The EFT scheme we have presented here for computing NN
scattering in perturbation theory appears to converge well and
preserve the desirable feature of the KSW scheme that at each order
the amplitude can be computed as a well defined set of Feynman
diagrams.  Unlike the KSW scheme, there is now a new dimensionful
parameter $\lambda$ which regulates the short distance tensor
interaction.  The manner with which we have performed this regulation
is certainly not unique, and we have shown that our results are not
particularly sensitive to the value of $\lambda$, and that over a wide
range for $\lambda$ the variation of the phaseshifts are comparable to
or smaller than higher order corrections in the EFT expansion.

On the other hand, we know that by taking $\lambda\to \infty$ we
recover the KSW expansion, which fails to converge above $p\sim
100\MeV$.  The parameter $\lambda$ apparently plays a role analogous
to the renormalization scale $\mu$ in perturbative QCD.  The scale
$\mu$ is unphysical, and a nonperturbative QCD calculation will not
depend on it; however, at any finite order in perturbation theory,
amplitudes do depend on $\mu$, and varying $\mu$ corresponds to
reordering the perturbative expansion.  Choosing $\mu$ appropriately
(e.g., via the BLM scale-setting prescription \cite{Brodsky:1982gc})
can optimize the perturbative expansion, while non-optimal choices for
$\mu$ lead to poor convergence.  Similarly, $\lambda$ is an unphysical
parameter, and varying $\lambda$ constitutes a reordering of the the
EFT expansion, with smaller $\lambda$ resulting in more of the pion
interaction being accounted for in the resummed contact interactions.
Taking $\lambda \simeq 750\MeV$ appears to optimize the expansion,
while choosing $\lambda=\infty$ yields the standard KSW expansion
which fails to converge at relatively low momenta.

It is not possible to directly compare our expansion with the Weinberg
expansion results at a given order, since the calculations are
arranged differently. For example, one-pion and two-pion exchange
appear at NLO and N${}^3$LO in the KSW expansion respectively, while
they appear at LO and NLO in the Weinberg expansion.  Nevertheless,
numerically our NNLO results compare favorably with the NLO Weinberg
expansion results in \cite{Epelbaum:1999dj}, with the exception of
$\epsilon_1$ which is comparable to LO.

The perturbative EFT described here provides a well defined
prescription for computing a number of additional processes to NNLO,
such as electromagnetic effects, including form factors, Compton
scattering, polarizabilities, and radiative capture, and it will be
interesting to compare such results with experiment to thoroughly
judge the efficacy of this method. An important outstanding issue is
the generalization of the power counting to account for the higher
partial waves. Evidently this poses a particularly difficult challenge
to EFT descriptions of nuclear forces and the perturbative formulation
described here is no exception. This issue will be addressed in future
work.

\section*{Acknowledgments}

\noindent I thank D.B.~Kaplan and A.~Vuorinen for a fruitful collaboration.
This work was supported by NSF CAREER Grant No. PHY-0645570.


\begin{thebibliography}{99}

\bibitem{Weinberg:1990rz}
  S.~Weinberg,
  Phys.\ Lett.\  B {\bf 251}, 288 (1990).

\bibitem{Weinberg:1991um}
  S.~Weinberg,
   ``Effective Chiral Lagrangians For Nucleon - Pion Interactions And Nuclear
  Nucl.\ Phys.\  B {\bf 363}, 3 (1991).

\bibitem{Weinberg:1992yk}
  S.~Weinberg,
  Phys.\ Lett.\  B {\bf 295}, 114 (1992)
  [arXiv:hep-ph/9209257].

\bibitem{Ordonez:1992xp}
  C.~Ordonez and U.~van Kolck,
  Phys.\ Lett.\  B {\bf 291}, 459 (1992).

\bibitem{Ordonez:1993tn}
  C.~Ordonez, L.~Ray and U.~van Kolck,
  Phys.\ Rev.\ Lett.\  {\bf 72}, 1982 (1994).

\bibitem{vanKolck:1994yi}
  U.~van Kolck,
  Phys.\ Rev.\  C {\bf 49}, 2932 (1994).

\bibitem{Ordonez:1995rz}
  C.~Ordonez, L.~Ray and U.~van Kolck,
  Phys.\ Rev.\  C {\bf 53}, 2086 (1996)
  [arXiv:hep-ph/9511380].

\bibitem{Friar:1998zt}
  J.~L.~Friar, D.~Huber and U.~van Kolck,
  Phys.\ Rev.\  C {\bf 59}, 53 (1999)
  [arXiv:nucl-th/9809065].

\bibitem{Rentmeester:1999vw}
  M.~C.~M.~Rentmeester, R.~G.~E.~Timmermans, J.~L.~Friar and J.~J.~de Swart,
  Phys.\ Rev.\ Lett.\  {\bf 82}, 4992 (1999)
  [arXiv:nucl-th/9901054].

\bibitem{Bernard:1996gq}
  V.~Bernard, N.~Kaiser and U.~G.~Meissner,
   ``Determination of the low-energy constants of the next-to-leading order
  Nucl.\ Phys.\  A {\bf 615}, 483 (1997)
  [arXiv:hep-ph/9611253].

\bibitem{Epelbaum:1999dj}
  E.~Epelbaum, W.~Gloeckle and U.~G.~Meissner,
   ``Nuclear forces from chiral Lagrangians using the method of unitary
  Nucl.\ Phys.\  A {\bf 671}, 295 (2000)
  [arXiv:nucl-th/9910064].

\bibitem{Epelbaum:2000mx}
  E.~Epelbaum, H.~Kamada, A.~Nogga, H.~Witala, W.~Gloeckle and U.~G.~Meissner,
  Phys.\ Rev.\ Lett.\  {\bf 86}, 4787 (2001)
  [arXiv:nucl-th/0007057].

\bibitem{Epelbaum:2002ji}
  E.~Epelbaum, A.~Nogga, W.~Gloeckle, H.~Kamada, U.~G.~Meissner and H.~Witala,
   ``Few nucleon systems with two-nucleon forces from chiral effective field
  Eur.\ Phys.\ J.\  A {\bf 15}, 543 (2002)
  [arXiv:nucl-th/0201064].

\bibitem{Epelbaum:2003xx}
  E.~Epelbaum, W.~Gloeckle and U.~G.~Meissner,
   ``Improving the convergence of the chiral expansion for nuclear forces.  II:
  Eur.\ Phys.\ J.\  A {\bf 19}, 401 (2004)
  [arXiv:nucl-th/0308010].

\bibitem{Epelbaum:2004fk}
  E.~Epelbaum, W.~Glockle and U.~G.~Meissner,
  Nucl.\ Phys.\  A {\bf 747}, 362 (2005)
  [arXiv:nucl-th/0405048].

\bibitem{Entem:2001cg}
  D.~R.~Entem and R.~Machleidt,
   ``Accurate nucleon nucleon potential based upon chiral perturbation
  Phys.\ Lett.\  B {\bf 524}, 93 (2002)
  [arXiv:nucl-th/0108057].

\bibitem{Entem:2002sf}
  D.~R.~Entem and R.~Machleidt,
  Phys.\ Rev.\  C {\bf 66}, 014002 (2002)
  [arXiv:nucl-th/0202039].

\bibitem{Entem:2003ft}
  D.~R.~Entem and R.~Machleidt,
   ``Accurate Charge-Dependent Nucleon-Nucleon Potential at Fourth Order of
  Phys.\ Rev.\  C {\bf 68}, 041001 (2003)
  [arXiv:nucl-th/0304018].

\bibitem{PavonValderrama:2005wv}
  M.~Pavon Valderrama and E.~R.~Arriola,
   ``Renormalization of NN Interaction with Chiral Two Pion Exchange Potential.
  Phys.\ Rev.\  C {\bf 74}, 054001 (2006)
  [arXiv:nucl-th/0506047].

\bibitem{PavonValderrama:2005uj}
  M.~Pavon Valderrama and E.~Ruiz Arriola,
   ``Renormalization of NN Interaction with Chiral Two Pion Exchange Potential.
  Phys.\ Rev.\  C {\bf 74}, 064004 (2006)
  [Erratum-ibid.\  C {\bf 75}, 059905 (2007)]
  [arXiv:nucl-th/0507075].

\bibitem{Kaplan:1996xu}
  D.~B.~Kaplan, M.~J.~Savage and M.~B.~Wise,
  Nucl.\ Phys.\  B {\bf 478}, 629 (1996)
  [arXiv:nucl-th/9605002].

\bibitem{Beane:2001bc}
  S.~R.~Beane, P.~F.~Bedaque, M.~J.~Savage and U.~van Kolck,
  Nucl.\ Phys.\  A {\bf 700}, 377 (2002)
  [arXiv:nucl-th/0104030].

\bibitem{Nogga:2005hy}
  A.~Nogga, R.~G.~E.~Timmermans and U.~van Kolck,
  Phys.\ Rev.\  C {\bf 72}, 054006 (2005)
  [arXiv:nucl-th/0506005].

\bibitem{Kaplan:1998we}
  D.~B.~Kaplan, M.~J.~Savage and M.~B.~Wise,
  Nucl.\ Phys.\  B {\bf 534}, 329 (1998)
  [arXiv:nucl-th/9802075].

\bibitem{Kaplan:1998tg}
  D.~B.~Kaplan, M.~J.~Savage and M.~B.~Wise,
  Phys.\ Lett.\  B {\bf 424}, 390 (1998)
  [arXiv:nucl-th/9801034].

\bibitem{Fleming:1999bs}
  S.~Fleming, T.~Mehen and I.~W.~Stewart,
  Phys.\ Rev.\  C {\bf 61}, 044005 (2000)
  [arXiv:nucl-th/9906056].

\bibitem{Fleming:1999ee}
  S.~Fleming, T.~Mehen and I.~W.~Stewart,
  Nucl.\ Phys.\  A {\bf 677}, 313 (2000)
  [arXiv:nucl-th/9911001].

\bibitem{Stoks:1993tb}
  V.~G.~J.~Stoks, R.~A.~M.~Kompl, M.~C.~M.~Rentmeester and J.~J.~de Swart,
   ``Partial wave analaysis of all nucleon-nucleon scattering data below
  Phys.\ Rev.\  C {\bf 48}, 792 (1993).

\bibitem{Beane:2008bt}
  S.~R.~Beane, D.~B.~Kaplan and A.~Vuorinen,
  Phys.\ Rev.\  C {\bf 80}, 011001 (2009)
  [arXiv:0812.3938 [nucl-th]].

\bibitem{Brodsky:1982gc}
  S.~J.~Brodsky, G.~P.~Lepage and P.~B.~Mackenzie,
   ``On The Elimination Of Scale Ambiguities In Perturbative Quantum
  Phys.\ Rev.\  D {\bf 28}, 228 (1983).



\end{thebibliography}
\end{document}